\documentclass[conference]{IEEEtran}

\usepackage{epstopdf}
\usepackage{amsmath} 
\usepackage{amsfonts} 
\usepackage{graphicx}
\usepackage{float}
\usepackage{url}
\usepackage{booktabs}
\usepackage{todonotes}

\usepackage{comment}
\usepackage[caption=false,font=scriptsize]{subfig}

\usepackage{tcolorbox}
\usepackage{multirow}
\usepackage{enumitem}
\tcbuselibrary{skins}

\newcommand{\idf}{{\tt Indexed DataFrame}}
\newcommand{\idfregular}{Indexed DataFrame}

\definecolor{BgBlue}{HTML}{336699}
\definecolor{BgGreen}{HTML}{339966}
\definecolor{BgYellow}{HTML}{664422}
\definecolor{BgOrange}{HTML}{994422}

\usepackage{listings}
\usepackage{color}

\definecolor{dkgreen}{rgb}{0,0.6,0}
\definecolor{gray}{rgb}{0.5,0.5,0.5}
\definecolor{mauve}{rgb}{0.58,0,0.82}

\begin{document}

\title{\huge In-Memory Indexed Caching for Distributed Data Processing}

\author{\IEEEauthorblockN{Alexandru Uta}
\IEEEauthorblockA{LIACS, Leiden University \\
\small a.uta@liacs.leidenuniv.nl}
\and
\IEEEauthorblockN{Bogdan Ghit}
\IEEEauthorblockA{Databricks \\
\small bogdan.ghit@databricks.com}
\and
\IEEEauthorblockN{Ankur Dave}
\IEEEauthorblockA{UC Berkeley \\
\small ankurd@eecs.berkeley.edu}
\and
\IEEEauthorblockN{Jan Rellermeyer}
\IEEEauthorblockA{TU Delft \\
\small j.s.rellermeyer@tudelft.nl}
\and
\IEEEauthorblockN{Peter Boncz}
\IEEEauthorblockA{CWI \\
\small p.boncz@cwi.nl}
}

\maketitle
\begin{abstract}

Powerful abstractions such as dataframes are only as efficient as their underlying runtime system. The de-facto distributed data processing framework, Apache Spark, is poorly suited for the modern cloud-based data-science workloads due to its outdated assumptions: static datasets analyzed using coarse-grained transformations.  
In this paper, we introduce the \idf{}, an in-memory cache that supports a dataframe abstraction which incorporates indexing capabilities to support fast lookup and join operations. Moreover, it supports appends with multi-version concurrency control.
We implement the \idfregular{} as a lightweight, standalone library which can be integrated with minimum effort in existing Spark programs. We analyze the performance of the \idfregular{} in cluster and cloud deployments with real-world datasets and benchmarks using both Apache Spark and Databricks Runtime. In our evaluation, we show that the \idfregular{} significantly speeds-up query execution when compared to a non-indexed dataframe, incurring modest memory overhead.

\end{abstract}
\section{Introduction}
\label{sec:intro}

The advent of data science has fundamentally changed our perception of how to gain insights, namely by adding what Jim Gray called a \emph{fourth paradigm} of science driven by data~\cite{hey2009fourth}. However, while pioneers like Gray envisioned relational database systems to become the engines of this new branch of science~\cite{barclay2000microsoft}, the tremendous momentum of data science has called for new systems to be developed, most importantly systems that are optimized for processing large amounts of unstructured or semi-structured data using clusters of machines. Due to the more agile and iterative nature of data science, those systems have departed from the idea of forcing data into a fixed schema for the purpose of giving the database system the chance to optimize common operations through query optimization and the use of indexes.     

Dataframes~\cite{mckinney2012python,koalas}, for instance, are modern-day data science abstractions similar to relational tables that enable users to express computations through SQL-like interfaces~\cite{armbrust2015spark} and execute those computations on distributed processing frameworks such as Dask~\cite{rocklin2015dask} or Apache Spark~\cite{zaharia2012resilient}. The high-level interfaces enabled by dataframes and SQL are very attractive to users as their programs can automatically trigger query optimization~\cite{armbrust2015spark} without manual tuning.

Even though dataframes are widely adopted by data scientists, they are only as efficient as their underlying runtime system can be. In practice, the performance can be underwhelming because systems like Spark have been designed around assumptions like the static nature of  data and justifies the reliance on coarse-grained transformations as the main processing paradigm but are, unfortunately, increasingly obsolete~\cite{vuppalapati2020building}. In the last decade new use cases for data science workloads emerged in which data can be processed through streaming interfaces~\cite{kreps2011kafka, armbrust2018structured} and data-lakes~\cite{armbrust2017databricks,potharaju2020helios}, which makes existing data processing pipelines to not necessarily run on static read-only files. The net result is an inefficient setup which is bottlenecked by network and IO bandwidth due to the reliance on shuffle and broadcast operations~\cite{uta2020big}.



Traditionally, data indexing has been a very effective way of minimizing the network overhead as it can significantly reduce the amount of data transferred by pre-filtering~\cite{DBLP:conf/sigir/DanzigANO91,DBLP:conf/ccgrid/NamS05,DBLP:conf/bigdataconf/DehneKRZZ13,DBLP:conf/sigmod/DiaconuFILMSVZ13,DBLP:conf/icde/GuptaHRU97}. 
However, supporting indexes on non-static datasets is difficult since write operations may cause consistency issues when scheduling tasks. This is particularly the case when using an external index as it as been suggested by prior work~\cite{ramnarayan2016snappydata}. 

In this paper, we present a novel approach that combines the best of the two worlds of relational database technology and data science systems. We show how embedding a write-enabled in-memory indexed cache into structures like Dataframes unlocks better performance for modern applications while seamlessly integrating into processing frameworks like Spark. 
The result of this effort, the \idfregular{}, enables low-latency joins and point look-ups in interactive workloads on data that is continuously changing and increasing in size. The \idfregular{} extends the space of use cases for Spark by efficiently supporting applications such as on-line threat detection and response~\cite{threatDetectionApple}, or real-time social network monitoring and dashboarding~\cite{erling2015ldbc}.

We show empirically that the \idfregular{} can be seamlessly integrated in Apache Spark frameworks deployed in clusters, but also in production-ready cloud-based environments such as the Databricks Runtime~\cite{dbr}. The main benefit of the integration with existing frameworks is the automatic access to the framework's scheduling and fault-tolerance mechanisms. 

The contributions of this work are:
\begin{enumerate}
\item We motivate the need of the \idfregular{} by showing the inefficiencies encountered in Spark when running typical data science queries (Section~\ref{sec:motivation}).
\item We present the design and implementation of the novel \idfregular{}, including the API we support to index Spark dataframes, the integration with Spark's Catalyst optimizer, and the underlying indexing data structure (Section~\ref{sec:design}). 
\item We evaluate the performance of our reference implementation of the \idfregular{} which is available open-source\footnote{\url{https://github.com/alexandru-uta/IndexedDF}}. 
We demonstrate the scalability of the \idfregular{}, as well as the performance improvement of the \idfregular{} of up to 20X in production environments using several real-world workloads and datasets (Section~\ref{sec:results}). 
\end{enumerate}

\section{A Case for Indexed Dataframes}\label{sec:motivation}

\begin{figure}[t]
\centering
\subfloat[Vanilla Spark.]{
  \includegraphics[width=1.6in, height=1.0in]{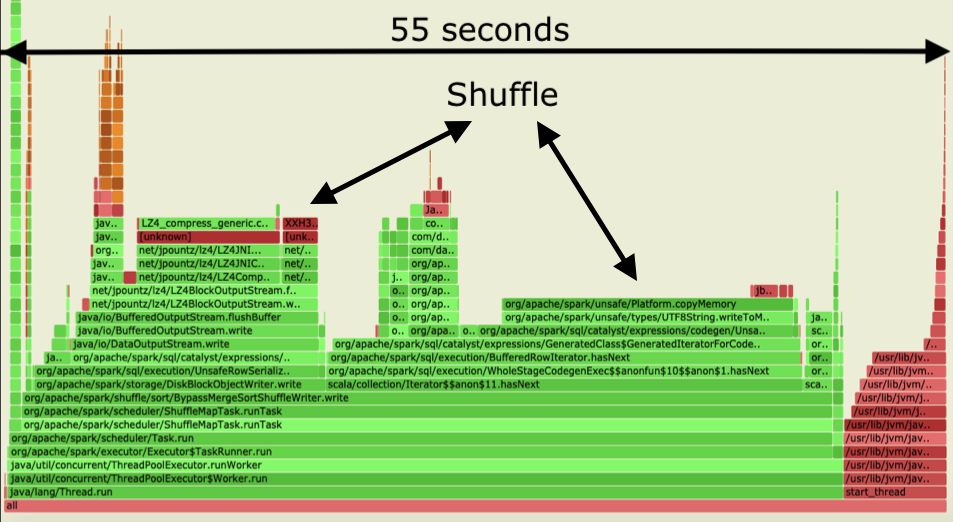}
  \label{fig:vanilla}
}
\subfloat[\idfregular{}.]{
  \includegraphics[width=1.6in, height=1.0in]{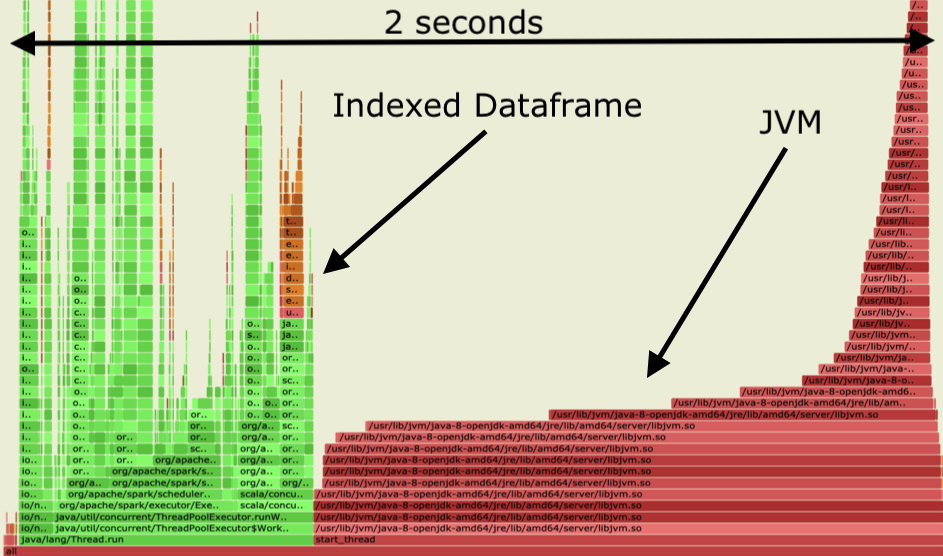}
  \label{fig:idf}
}
\caption{Flame-graphs for 5 consecutive runs of a \emph{join} operation on a Spark worker: Vanilla Spark (left) and the \idf{} (right) running on the Databricks Runtime. The workload is 5 consecutive join operations. The red part at the right of each subfigure represents JVM load, while the green parts represent actual Spark computations.}
\label{fig:flamegraphs}
\end{figure}

Indexing is a well-known technique employed in traditional database systems to speed-up the access to the data source. In this section, we identify the main reasons why indexing is attractive for modern data science workloads and we explain why adopting them in data analytics engines such as Spark is challenging. 
In social-network graph processing~\cite{erling2015ldbc} and cyber-security threat detection~\cite{threatDetectionApple} many similar operations are performed continuously, while the datasets are constantly growing, often through fine-grained (individual records or small sets of) appends. Data processing frameworks such as Spark are not designed for offering sufficient performance to support such application patterns.

These applications make extensive use of \emph{point lookups} and \emph{join} operations. The former relate to quickly finding certain elements in a large data collection. The latter relate to combining information in multiple columns of two or more dataframes. Furthermore, both aforementioned applications could benefit from fine-grained \emph{appends}: in the graph processing use-case, new links in social network graphs are formed continuously, while in real-time threat detection and security monitoring, network connections are incoming in high-volumes, and need to be analyzed in interactive time.

Hybrid transaction/analytical processing (HTAP) systems such as Druid~\cite{yahoodruid}, Splice Machine~\cite{splicemachine}, or SnappyData~\cite{ramnarayan2016snappydata} try to solve such problems by reconciling both OLAP and OLTP workloads. Designing such capabilities require a rethinking of the underlying runtime system such that it efficiently supports both types of workloads, or augmenting it with another system, leading to maintenance difficulties. 

Being designed with immutability in mind, Spark only supports coarse-grained data transformation and not fine-grained updates/appends. To support updates, Spark needs to be integrated with external storage such as Cassandra~\cite{lakshman2010cassandra} or Delta Tables~\cite{armbrust2017databricks} or the Azure implementation of Data Lakes~\cite{potharaju2020helios,potharajuhyperspace}. For running queries on fresh data, \emph{even if only few records have been added}, Spark requires reloading the complete dataset from the external data store after a write. Without supporting fine-grained appends in-place, reloading data from external data sources is an expensive operation, \emph{which highly limits interactive response times}.

Spark is generally inefficient for point lookups and joins. Without additional data structures or partitioning, point lookups in Spark are linear in time to the number of entries. Joins are even more complex due to Spark's distributed nature: data is either sorted and then merged (i.e., Sort-Merge Join~\cite{graefe1994sort}), or hash-tables are being built for one of the dataframes, these are then broadcast and probed locally against all entries of the other dataframe (i.e., BroadcastHash Join~\cite{dewitt1985multiprocessor}). In data science operational pipelines, these operations are not run only once, but continuously. Therefore, performing $O(n)$ operations for lookups, building hash-tables and shuffling data around for every run is inefficient. Implementing an \emph{index} next to the data helps both these operations: point lookups now become worst-case logarithmic time, while for joins the index acts as a pre-built hash-table. 

To show evidence for this, we performed 5 joins operations in a sequence on a 7$GB$ Broconn~\cite{threatDetectionApple} table, joining it with a small random sampled subset of itself, of less than 10$MB$. We ran these join operations on the Databricks Runtime on four \texttt{i3.xlarge} virtual machine instances.
Figure~\ref{fig:flamegraphs} shows the performance breakdown for these two executions. The regular Spark implementation for join operations needs to perform the same networked operations and hash-table building for each join execution. For the \idfregular{}, the index is computed only once, and its overhead can be amortized over many executions of the indexed operations.

We have argued for and gave empirical evidence that supports the addition of indexes in Spark. In the following sections, we provide the in-depth design and performance evaluation of the \idfregular{}.

\section{\idfregular{} Design}
\label{sec:design}

\begin{lstlisting}[caption=The Indexed Data Frame API., label=lst:indexed_api, float, belowskip=-0.99\baselineskip]
// creating an index
var indexedDF = regularDF.createIndex(colNo)
// caching the indexed dataframe
var indexedDF = indexedDF.cache()
// key lookup returns a dataframe 
val lookupKey = 1234
val resultDataFrame = indexedDF.getRows(lookupKey)
// appending all the rows of a regular dataframe
val newIndexedDF = indexedDF.appendRows(aRegularDF)
// index-powered, efficient join
val result = indexedDF.join(regularDF, indexedDF.col("c1") === regularDF.col("c2"))
\end{lstlisting}

We propose the \idf{}, a data abstraction that we can use to manipulate cached indexed datasets. 
We present the API and the main design and implementation details. The \idfregular{} is built as an extension library which integrates with the Dataframe API supported by Spark SQL~\cite{armbrust2015spark}, and can be added to existing Spark programs. 

\subsection{Programming Interface} 
To address the requirements of a wide spectrum of large-scale data science applications, we designed the \idfregular{} to support the following: \texttt{create index}, {\tt cache index}, {\tt point lookups}, {\tt append rows}, and {\tt indexed joins}. The corresponding Scala API is presented in Listing~\ref{lst:indexed_api}.
In our current implementation the index supports any type of column, but for good performance, we recommend using only primitive column types (e.g., (un)signed 32/64-bit integers, floating point numbers). 

\begin{figure}[t]
\centering
\includegraphics[width=0.72\columnwidth]{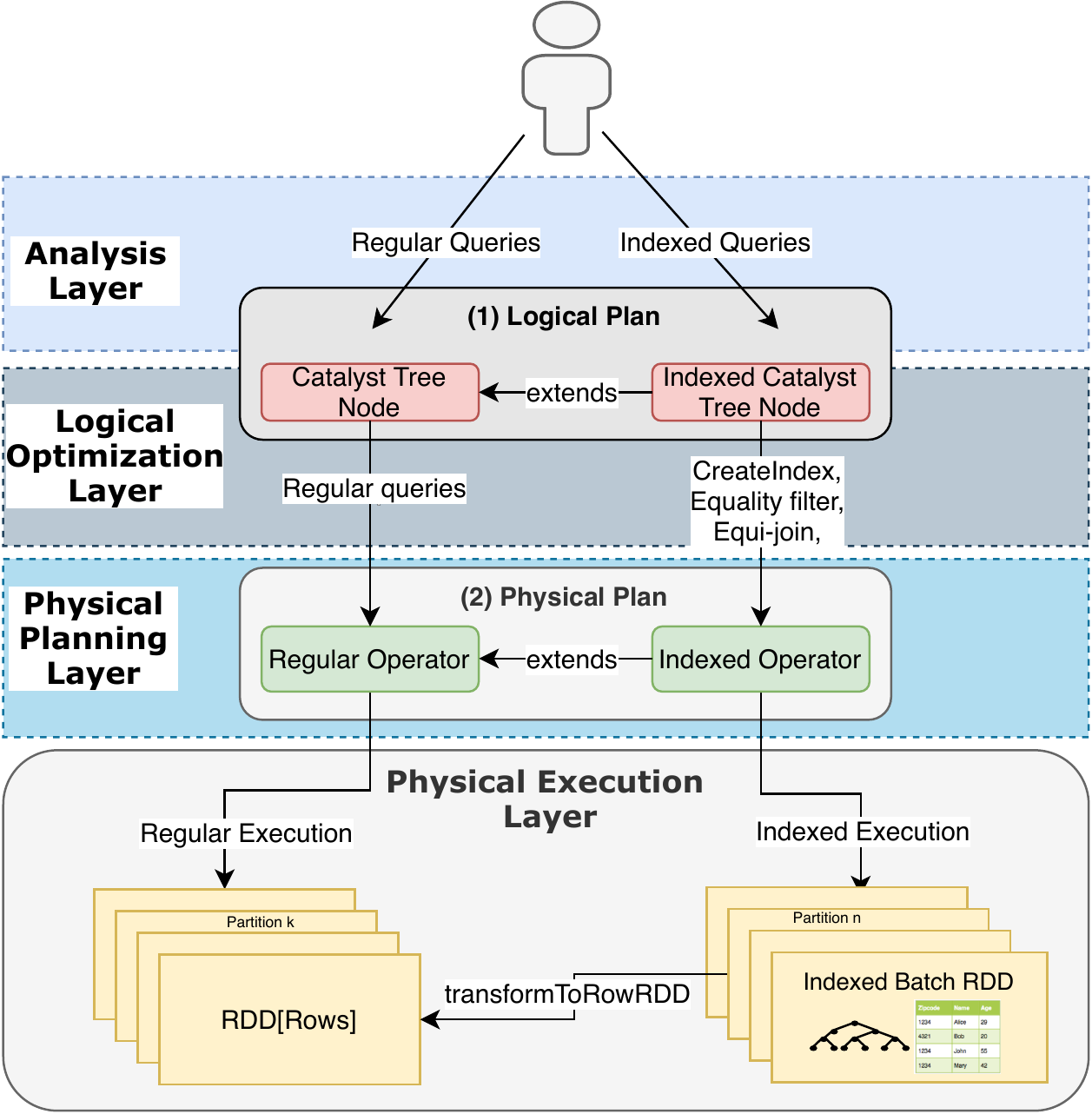}
\caption{\idfregular{} logical flow. Users write SQL queries or use the Dataframe API. Catalyst rules determine whether the queries are regular or indexed. If regular, they follow the regular execution. If indexed, special rules, and optimization strategies are applied such that indexed execution is triggered. An Indexed Batch RDD can always fall back to a regular Spark Row RDD to trigger regular execution on top of the \idfregular{}.}
\label{fig:high_level_design}
\vspace*{-0.4cm}
\end{figure}

As we want to store the \idfregular{} in-memory on the Spark executors, instantiating the index should be immediately followed by a caching operation. Furthermore, the {\it append rows} operation can be performed both in a fine-grained and a batch-oriented mode by organizing the rows we need to append as a regular Spark Dataframe. In this way, users can append with low latency small amounts of rows, or batch multiple updates in a larger Dataframe. When users need to lookup the rows associated with a certain key, our library returns a (smaller) Dataframe containing the required rows. In the case of \emph{join} operations, if any of the sides of the relation are indexed, our implementation of the \idf{} triggers an indexed join operation. The result is a regular Spark dataframe. Evidently, in case of the \emph{indexed join}, the indexed relation is always the build side (as it is actually pre-built due to the index), while the probe side is the non-indexed relation.

\subsection{Integration with Catalyst}
Figure~\ref{fig:high_level_design} shows the architecture of the \idf{} and its integration with the Catalyst optimizer in Spark. To add indexed operations to the regular Spark SQL and the Dataframe API without modifying the Spark source code we employ Scala \emph{implicit conversions}. In this way we can add our methods to the Dataframe class, while leveraging the full capabilities of the Catalyst~\cite{armbrust2015spark} query optimizer. Our library includes \emph{optimization rules} that make regular Spark SQL queries aware of our custom indexed operations. 

In Spark SQL, queries have abstract representations called query plans. These are converted through a sequence of transformations into optimized plans that finally execute on the cluster. Catalyst translates queries into logical plans that provide high-level representations of each operator without defining how to perform the computation. Optimization rules transform the logical plan into a physical plan with specific instructions on how to execute the query. 

Through our library, we use the extensibility of Catalyst to add index-aware optimization rules. These translate the indexed logical operators into physical operators. These rules ensure that the appropriate look-up functions are called for each indexed or basic logical operator and ensure that the \idfregular{} operations are always triggered when executing queries on indexed data. Similarly, for queries on non-indexed dataframes we fall back to the default Spark behavior.    

\subsection{The Indexed Batch RDD}~\label{sec:rdd}
Spark datasets are typically partitioned across multiple nodes so that the framework can divide jobs into multiple tasks that can be executed in parallel on multiple \emph{executors}. The Dataframe API can perform relational operations on Spark's built-in distributed collections, i.e., the {\sc RDD}s~\cite{zaharia2012resilient}. The \idf{} operates in a similar way by partitioning data across multiple executors, but requires a custom RDD implementation, the \textit{Indexed Batch RDD}, to make use of indexed operations.

\begin{figure}[t]
\centering
\includegraphics[width=0.72\columnwidth]{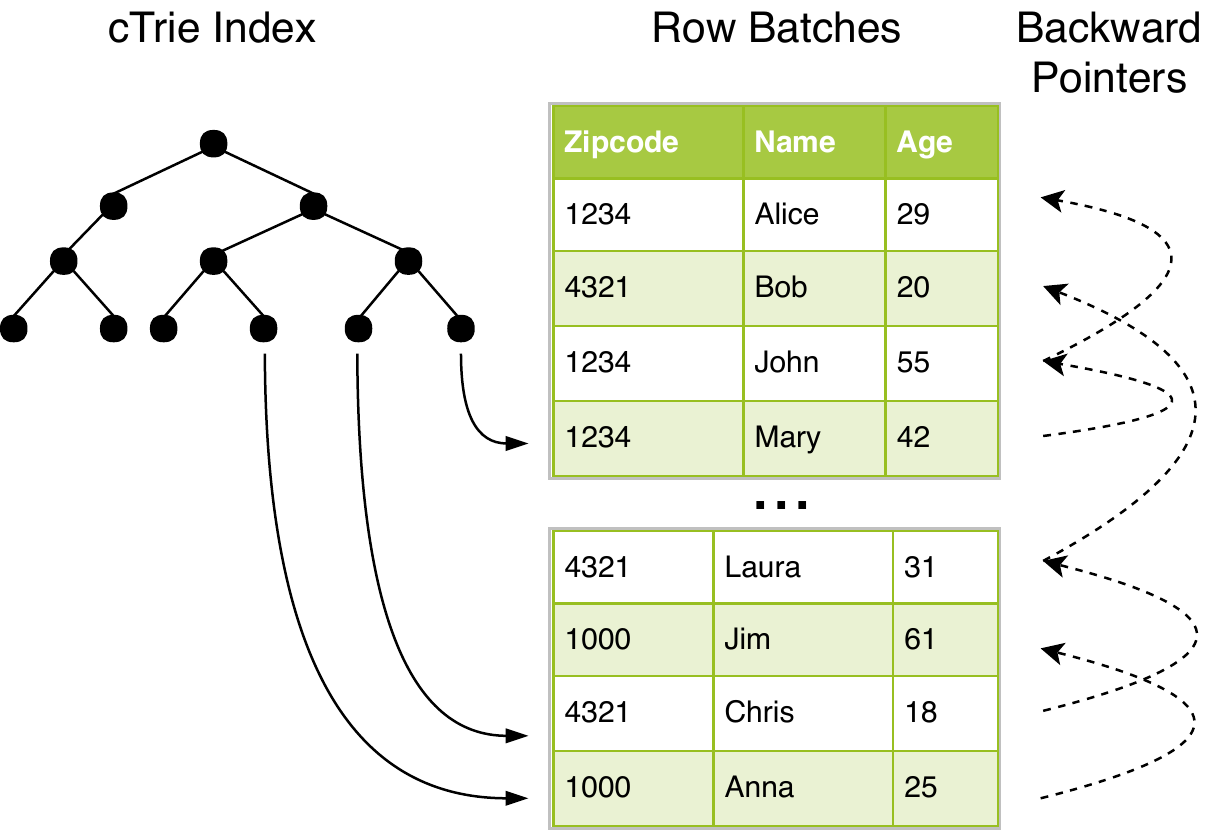}
\caption{\idfregular{} internal design: (1) cTrie for storing pointers to last row containing the key; (2) collection of row batches storing the data; (3) backward pointers for rows with equal keys.}
\label{fig:low_level_design}
\vspace*{-0.5cm}
\end{figure}

Figure~\ref{fig:low_level_design} depicts the design of the {\it Indexed Batch RDD}. Our implementation stores data \emph{in-memory}. This decision was made to optimize for performance but without loss of generality; the representation could easily extend to store data out-of-core, for example in SSD or NVMe devices for different tradeoffs. Each partition is composed of three data structures: 
(1) a \emph{cTrie}~\cite{prokopec2012concurrent}, which represents the index,
(2) a set of \emph{row batches}\footnote{In our prototype we store data in row-wise format in the \emph{Indexed Batch RDD}. However, this could seamlessly be changed to columnar formats. The decision is based on the type of workload the user needs to support.}, which store the tabular data, and
(3) a set of \emph{backward pointers}, which are used to crawl the partition for rows that are indexed on the same key. 

\textbf{Design.} The cTrie is a concurrent hash-trie, which can employ thread-safe, lock-free inserts, deletes, and lookups. Furthermore, the cTrie can perform lock-free, atomic snapshotting in constant time. The cTrie snapshotting is similar to a \emph{persistent data structure}~\cite{driscoll1989making}. Because a new snapshot version shares most of the state with the parent object, we only need to store the actual modifications resulted from appends. The cTrie requires minimum overhead when creating new \idfregular{}s by taking snapshots on writes.

\textbf{Non-unique Keys.} The cTrie stores a pointer to the latest appended row associated with a key. If there are multiple rows associated with a key, the backward pointer data structure consists of linked lists, one per unique key. This data structure can be used to traverse the list of rows associated with the key. The row batches are collections of binary, unsafe arrays (e.g., of 4MB in size), each storing a number of rows determined by the row and batch sizes. The pointers stored both in the cTrie and in the backward pointer data structure are packed in dense 64-bit integers, each containing the row batch number, an offset within a row batch, and the size of the previous row indexed on the same key. 

\textbf{Maximum Size.} In our experiments we use indexed partitions with rows that may have up to $1$~KB and $2^{31}$ row batches, each of which may have up to $4$~MB. Thus, our setup enables $4 \times 2^{31}$ MB data per core, sufficient for modern server configurations. Transformations within a partition are sequentially executed on a single core. As a rule-of-thumb, Spark deployments should be configured with 1 to 4 partitions per core\footnote{\url{https://spark.apache.org/docs/latest/tuning.html}}. Both the batch and row sizes are configurable.

\textbf{Scheduling Physical Operators.} To implement the \emph{indexed} operations efficiently in Spark, we employ a \emph{hash partitioning} scheme on the indexed key and shuffle operations to transfer the data to their indexed partitions. This section presents the main \idf{} operations.

\textbf{Index Creation, Append.} The operation of the \idfregular{} on a single partition is similar both for index creation and for appends. The \idfregular{} is \emph{hash} partitioned on the indexed column. This ensures a better load balancing when the key ranges are not known a-priori. When an index is created on a regular Dataframe, its rows are shuffled based on the hash partitioning scheme to their respective \idfregular{} partitions. First, each row is inserted in the row batch responsible for its key. If a row with a similar key was already inserted in the partition, the cTrie entry for the key is updated to point to the newly added row, while the backward pointer of the newly added row is created to point to the previous row. 

\textbf{Lookup.} A lookup operation is scheduled on the Spark partition responsible for holding that key based on the hash partitioning scheme. The lookup is then performed locally, and the set of resulting rows is returned to the user as a regular Dataframe. A local lookup consists of a search in the cTrie, followed by a traversal of the backward pointers if multiple rows share the same key.

\textbf{Indexed Join.} To join a \idfregular{} and a (regular) Dataframe, the rows of the latter are shuffled according to the hash partitioning scheme of the former. As the \emph{build} side is already created in the form of the index, the \emph{probes} are made locally from the shuffled rows. If the Dataframe size is small enough to be broadcasted efficiently, we fall back to a broadcast-based join instead of a shuffle.

\begin{lstlisting}[caption=Indexed DataFrame divergence.,label=lst:indexed_divergence, float, floatplacement=b,belowskip=-0.3\baselineskip]
// dataframes A and B share the same parent
indexedDF_A = indexedDF.append(appendDF1)
indexedDF_B = indexedDF.append(appendDF2)
// divergent and materialized in reverse order
indexedDF_B.collect()
indexedDF_A.collect()
\end{lstlisting}

\subsection{Fault-tolerance and Consistency}
\label{sec:ft}

Spark fault-tolerance is achieved via task recomputation based on its lineage graph. To achieve fault-tolerance for the \idf{}, we use the recomputation mechanism provided by Spark in the following way. All \idfregular{} operations, except \emph{append}, are able to be replayed using the lineage graph without any additional considerations. For the append operation to be re-generated properly, we rely, similarly to Spark Structured Streaming~\cite{armbrust2018structured}, on either a replayable data source, such as Apache Kafka~\cite{kreps2011kafka} or a persistent (distributed) file system, such as HDFS~\cite{shvachko2010hadoop}.    

Maintaining consistency when supporting data appends is non-trivial. We distinguish special use-cases: supporting straggler nodes, bad scheduler decisions, or even failed nodes. In either of these cases, a Spark task can be scheduled on a node that does not have locally the required indexed partition. Spark is optimized for achieving data locality, so every task is prioritized to run on a node that contains its input data. In case locality cannot be achieved within a configurable timeout, tasks can be scheduled on remote nodes. In such a case, in practice we end up with multiple copies of the same data. If this piece of data is obtained through multiple transformations, these are replayed on the new node.

In the case of \idfregular{}, this mechanism works similarly. This means that all former operations must be replayed locally, which is a safe operation but results in two \emph{identical} copies of the same data. While in regular Spark operations this is a non-issue, since data is static,
a new \emph{append} operation means that the two partitions are not identical anymore. Hence, the older partition cannot be used for future tasks, because it is stale. To properly handle these situations, on each \emph{append} on an \idfregular{}, the underlying custom RDD data structure (i.e., Indexed Batch RDD) increments a \emph{version number}. The version number aids the scheduler not to send tasks to \emph{stale} partitions, ensuring \emph{consistent} operation.

\begin{table}[t]
\centering
\caption{The hardware configuration used in our experiments.}
\vspace*{-0.3cm}
\resizebox{0.98\columnwidth}{!}{
\begin{tabular}{@{}llllll@{}}
\toprule
\textbf{Hardware}                                         & \textbf{Type}                                               & \textbf{\#Cores} & \textbf{Memory} & \textbf{Network}                                          & \textbf{Disk} \\ \midrule
\begin{tabular}[c]{@{}l@{}}Private\\ Cluster\end{tabular} & \begin{tabular}[c]{@{}l@{}}Intel \\ E5-2630-v3\end{tabular} & 16               & 64 GB           & \begin{tabular}[c]{@{}l@{}}FDR \\ InfiniBand\end{tabular} & SSD           \\ \midrule
Amazon EC2                                                & \begin{tabular}[c]{@{}l@{}}i3.xlarge \& \\ i3.8xlarge\end{tabular}                                                 & 4 \& 16               & 30 \& 122 GB          & 10 Gbps                                                   & SSD           \\ \bottomrule
\end{tabular}
}
\label{tab:hardware}
\vspace*{-0.45cm}
\end{table}

\subsection{Divergence and Multi-Versioning}

Because the \emph{append} operation returns a new version of the \idfregular{}, we can reach the situation depicted in Listing~\ref{lst:indexed_divergence}. Here, two successive appends on a parent data frame create two divergent children dataframes. However, the append is only triggered when the newly created \idfregular{} is materialized. Theoretically, we could make the parent dataframe read-only once an append is performed, to guard against such situations. In practice, in case the children dataframes are materialized in reverse order, it is not trivial to decide which append should be permitted.

To permit both appends, a pragmatic solution would be to employ a \emph{copy-on-write} mechanism, such that divergent dataframes could co-exist. However, this incurs large performance penalties (i.e., full data copies) and storage overheads (i.e., keeping multiple copies of the same data). An efficient solution is to employ a \emph{persistent data structure} scheme, where the children dataframes share the parent data and only store the \emph{deltas}. This is achievable due to the cTrie index capability: whenever a snapshot is triggered, the newly created copy shares the initial state with no memory overhead and only stores differences to the previous version. In case of the row batches, this is achieved with a similar scheme: we use a secondary cTrie that stores pointers to the row batches of the \idfregular{} partitions. In this way, the \idfregular{} supports divergent appends with \emph{minimum storage and performance overheads}.

\begin{table}[t]
\caption{The real-world and synthetic datasets and the queries we use to evaluate the performance of the \idfregular{}.}
\vspace*{-0.2cm}
\resizebox{0.98\columnwidth}{!}{
\begin{tabular}{lllll}
\toprule
\multicolumn{1}{c}{\textbf{Dataset}} & \multicolumn{1}{c}{\textbf{Experiment}} & \multicolumn{1}{c}{\textbf{Query}} & \multicolumn{1}{c}{\textbf{Query Desc.}} & \begin{tabular}[c]{@{}l@{}}\textbf{Index}\\\textbf{Column}\end{tabular} \\ \midrule

SNB (SF-1000) & \S\ref{subsec:deployment},\ref{subsec:scalability},\ref{subsec:micro} & Join & \begin{tabular}[c]{@{}l@{}}join edges with vertices ON\\edge\_source\end{tabular}  & integer \\ \midrule

SNB (SF-300) & \S\ref{subsec:realworld} & SQ1-SQ7 & online\footnote{\url{http://ldbc.github.io/ldbc_snb_docs/ldbc-snb-specification.pdf}} & various \\ \midrule
\multirow{8}{*}{\begin{tabular}[c]{@{}l@{}}US Flights\\(120 GB)\end{tabular}} & \S\ref{subsec:realworld} & Q1 & join flights with planes on tailNum & string \\
 & \S\ref{subsec:realworld} & Q2 & select * where tailNum = x & string \\
 & \S\ref{subsec:realworld} & Q3 & \begin{tabular}[c]{@{}l@{}}join flights with selected flights table\\ (flightNum \textless 200)\end{tabular} & integer \\
 & \S\ref{subsec:realworld} & Q4 & \begin{tabular}[c]{@{}l@{}}join flights with selected flights table\\ (flightNum \textless 400)\end{tabular} & integer \\
 & \S\ref{subsec:realworld} & Q5 & point query (10 matches) & integer \\ 
& \S\ref{subsec:realworld}  & Q6 & point query (100 matches) & integer \\
& \S\ref{subsec:realworld}  & Q7 & point query (1000 matches) & integer \\ \bottomrule
\begin{tabular}[c]{@{}l@{}}TPC-DS \\ (SF-1,10,100,1000)\end{tabular} & \S\ref{subsec:realworld} & Join & \begin{tabular}[c]{@{}l@{}}store\_sales JOIN date\_dim ON\\ ss\_sold\_date\_sk\end{tabular} & integer \\ \bottomrule
\end{tabular}
}
\label{tab:queries_description}
\vspace*{-0.45cm}
\end{table}

\subsection{Implementation}

The \idfregular{} was implemented in Scala and Java and is available as a standalone open-source \textit{sbt} project. Users can bundle the generated jar binary in their applications and simply have Spark make use of the Catalyst optimizations and strategies that trigger indexed operations. After a call to \emph{createIndex} is made (i.e., the only modification a programmer has to make to a Spark program to use the \idfregular{}), all the functionality we implemented is triggered automatically if the optimizer rules we provided determine that the index can be used. These include Spark SQL code and dataframe code.

The \idfregular{} allocates memory for the rows it stores in \textit{unsafe} off-heap memory that is not managed by the JVM, thus not triggering inefficient GC pauses. The only JVM-managed memory of the \idfregular{} is the cTrie, which is a scala-native implementation. However, as we show in Section~\ref{sec:results}, this is sufficiently effective in maintaining an index that achieves interactive query time. 



\section{\idfregular{} Evaluation}
\label{sec:results}

We take an experimental approach to evaluate performance aspects of the \idfregular{}. To this end, we seek to answer the following questions:

\begin{enumerate}[leftmargin=*]
\item[Q1] \textit{What (environment) parameters and variables is the \idfregular{} sensitive to?} 

\item[Q2] \textit{How does the \idfregular{} scale with the number of cluster machines, numbers of cores, and problem size?} 

\item[Q3] \textit{How does the \idfregular{} perform on typical Spark SQL operations (e.g., filter, projection, aggregation), and what are its overheads?} 

\item[Q4] \textit{How does the \idfregular{} perform on real-world applications and how does it integrate into production environments?} 

\end{enumerate}

\begin{table}[t]
\centering
\tiny
\caption{The size of each probe and build relation in different join operations used in our experiments.}
\vspace*{-0.2cm}
\resizebox{0.98\columnwidth}{!}{
\begin{tabular}{@{}lccc@{}}
\toprule
\multicolumn{1}{c}{\textbf{Join Scale}} & \multicolumn{1}{c}{\textbf{\begin{tabular}[c]{@{}c@{}}Probe Side\\ (rows)\end{tabular}}} & \multicolumn{1}{c}{\textbf{\begin{tabular}[c]{@{}c@{}}Indexed or Build\\ Side (rows)\end{tabular}}} & \multicolumn{1}{c}{\textbf{\begin{tabular}[c]{@{}c@{}}Result Size\\ (rows)\end{tabular}}} \\ \midrule
S          & 10K                                                                                      & 1B                                                                                          & 1.5M                                                                                          \\
M                                       & 100K                                                                                     & 1B                                                                                          &  14M                                                                                         \\
L                                      & 1M                                                                                       & 1B                                                                                          &   110M                                                                                        \\
XL                                       & 10M                                                                                      & 1B                                                                                          & 1B                                                                                          \\ \bottomrule
\end{tabular}
}
\label{tab:join_sizes}
\end{table}

\begin{figure}[t]
    \centering
    \includegraphics[width=0.75\columnwidth]{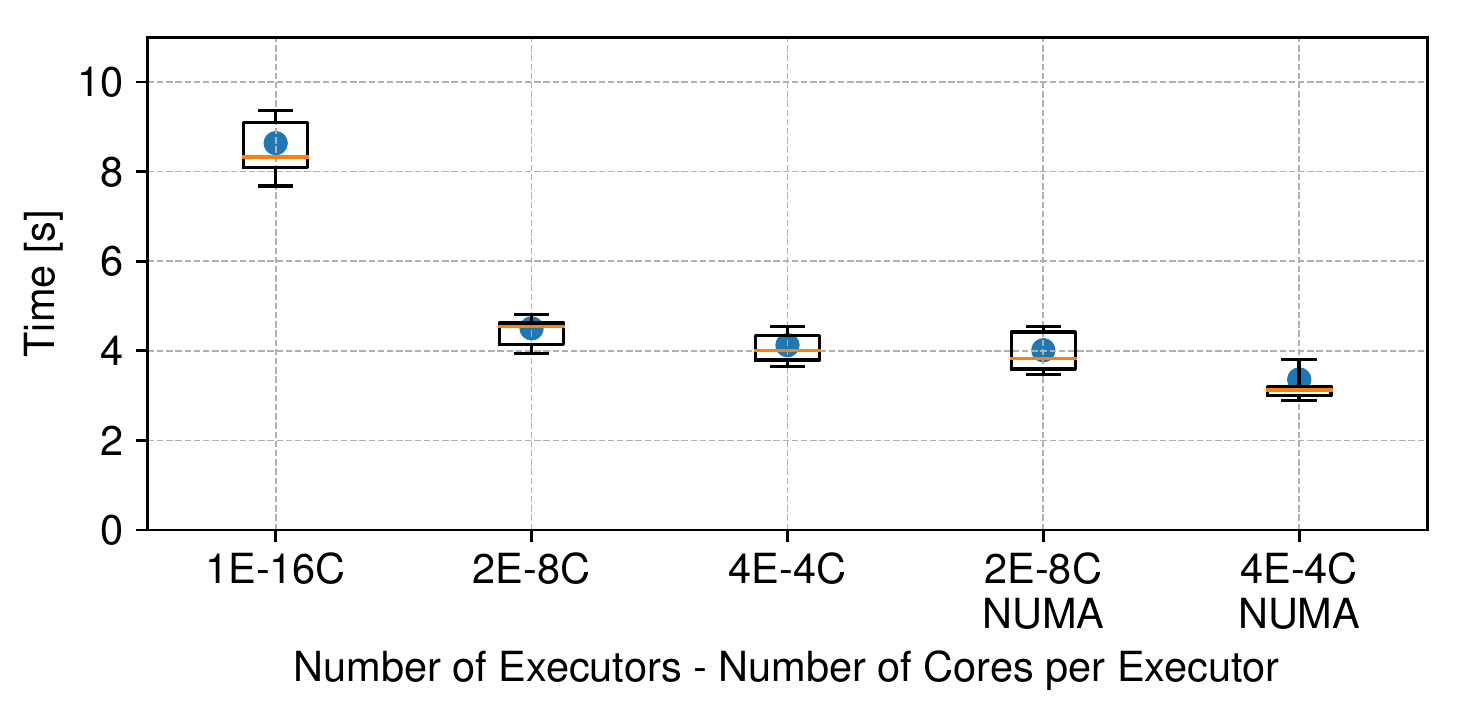}
    \vspace*{-0.25cm}
    \caption{Deployment of the \idf{} on a dual-socket NUMA machine. Number of Spark executors vs. cores per executor vs. pinning executors to NUMA domains. Data is presented as IQR boxplots, with whiskers representing minimum and maximum, while blue dot represents the mean performance.}
    \label{fig:deployment_analysis}
    \vspace*{-0.4cm}
\end{figure}

\subsection{Experimental Setup}
We evaluate the \idfregular{} through experiments on both a private non-virtualized cluster and a production system deployed on a public cloud. We present the cluster and cloud configurations and the SQL benchmarks that we use to assess the performance of the \idfregular{}.

\textbf{Cluster Setup.} 
To run our experiments we have deployed Spark 2.3.0 with the \idfregular{} on a private non-virtualized cluster~with the datasets stored in an HDFS instance with version 2.7.0. We co-located the HDFS instance with Spark on the same machines.
Furthermore, we incorporated the \idfregular{} with clusters provisioned from the Amazon EC2 public cloud with the Databricks Runtime~\cite{dbr} system with the datasets stored on S3. 
For each experiment, we report averages of performance metrics over many runs, adhering to modern standards of performance reproducibility~\cite{uta2020big,maricq2018taming}. 
The \idfregular{} is an in-memory table, thus our performance baseline is the default in-memory (columnar) caching mechanism provided by Spark.

\begin{figure}[t]
    \centering
    \includegraphics[width=0.85\columnwidth]{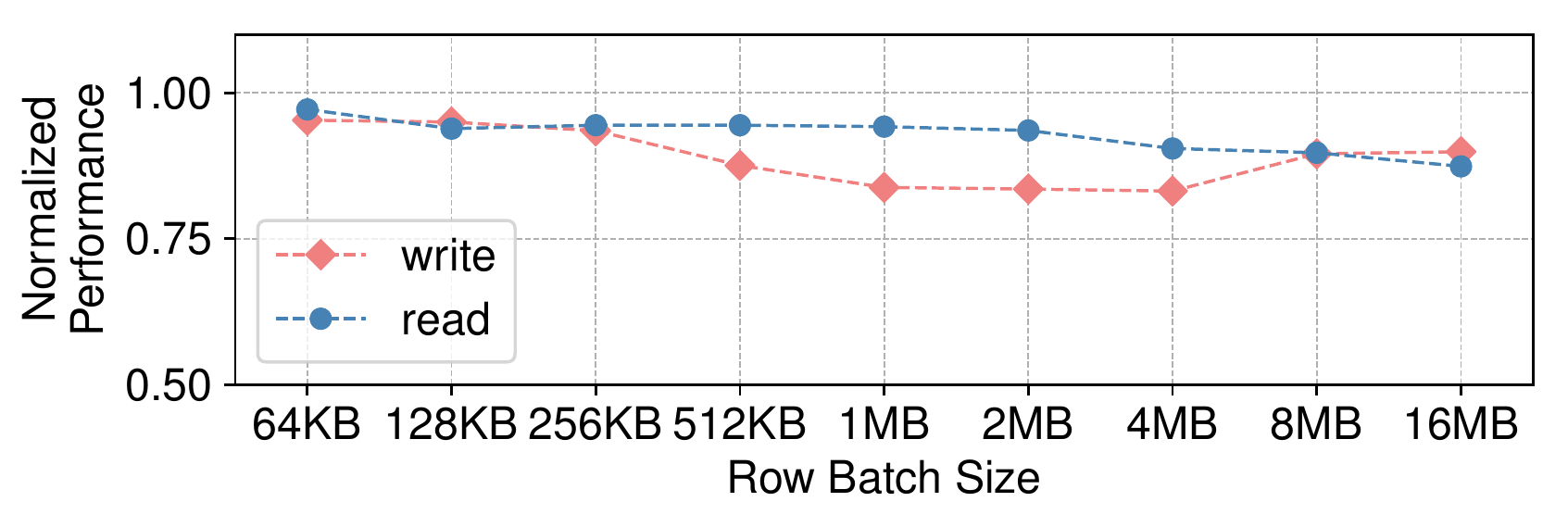}
    \vspace*{-0.25cm}
    \caption{Read and write performance for the \idfregular{} with varying row batch sizes. Results are normalized by the performance of a deployment where the batch size is 4KB (OS page size). Vertical axis does not start at 0 to improve visibility.}
    \label{fig:bs_sensitivity}
    \vspace*{-0.4cm}
\end{figure}

\textbf{Workloads and Datasets.} We use two types of queries in our experiments: lookups and joins. We run these against synthetic and real-world datasets which we summarize in Table~\ref{tab:queries_description}. The Social Network Benchmark (SNB)~\cite{erling2015ldbc} is designed to mimic typical social network structure and behavior. This benchmark was developed for evaluating data analytics applications on updatable graphs. The benchmark generates a social network with power-law structure, similar to Facebook. The benchmark also includes a number of queries to explore the graph.
The dataset consists of \emph{edge} and \emph{vertex} tables (jointly over 33\,GB), each of which stores various attributes of users in the network. 

TPC-DS~\cite{nambiar2006making} is the de-facto dataset for assessing the performance of large-scale analytics frameworks. We use various scale factors from 1 to 1000 to assess the performance of indexed joins.

Finally, we use a real-world dataset released by the US Department of Transportation which tracks the performance of domestic flights operated by large air carriers~\cite{flights}. We use two tables from this dataset, a flights table of 120\,GB and a planes table of 420\,KB. The \texttt{flights} table records the details of each flight that is identified by a unique number (\texttt{flightNum}) including the plane identifier, departure and arrival dates, delays, origin, and destination. 

\begin{figure}[t]
\centering
\subfloat[]{
  \includegraphics[width=0.8\columnwidth]{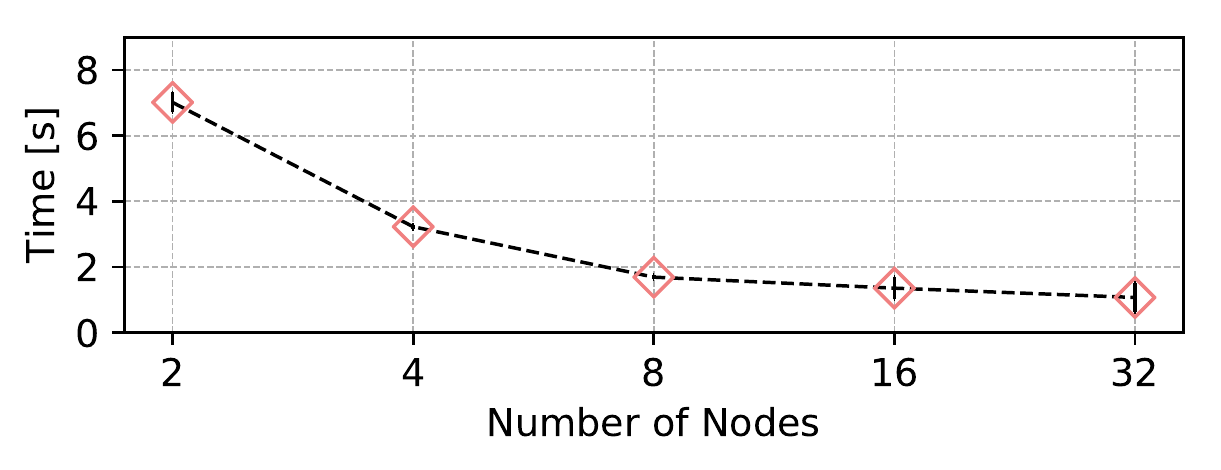}
    \label{fig:horizscal}
   
}
\\
 \vspace*{-0.9cm}
\subfloat{
  \includegraphics[width=0.8\columnwidth]{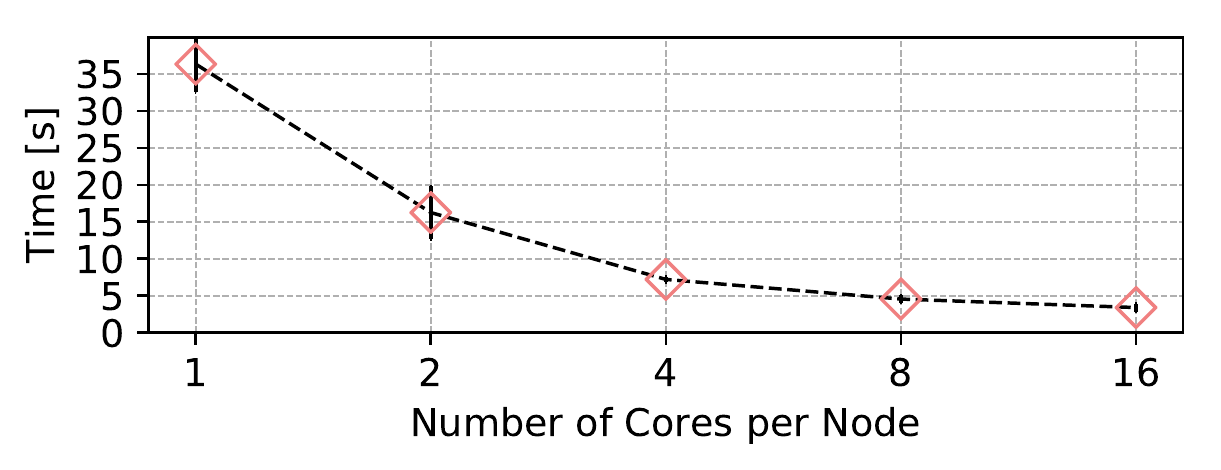}
    \label{fig:vertscal}
    
}
\vspace*{-0.45cm}
\caption{The scalability properties of the \idf{}: horizontal scalability (top) and vertical scalability (bottom). The points are average runtimes over 10 repetitions and the whiskers standard deviations.}
\label{fig:scalability}
\vspace*{-0.4cm}
\end{figure}

\subsection{Sensitivity Analysis}\label{subsec:deployment}

We show an investigation on what parameters make a difference when deploying and running the \idfregular{}.

{\bf Modern Servers.} Memory latency and bandwidth can be a significant bottleneck in large-scale data processing, especially when the execution platform is a complex NUMA architecture.  
Spark is JVM-based and thus agnostic to the underlying physical hardware of the machines. To deploy a Spark cluster, administrators need to decide how to slice the resources of the worker machines between multiple executors. Standard deployments are usually based on rules-of-thumb such as \emph{one executor per machine} or \emph{$n$ cores per executor}. Optimal allocation is not investigated in prior work focused on performance~\cite{DBLP:conf/nsdi/OusterhoutRRSC15,DBLP:conf/hpdc/ChaimovMCIIS16}.
Recent research~\cite{DBLP:conf/ispass/ChibaO16,DBLP:conf/bigdataservice/BaigAPC18} conducted on IBM Power8 clusters shows evidence that Spark is indeed sensitive to {\sc numa} allocations.


{\bf Impact of NUMA.} We assess the impact of {\sc numa} machines on the performance of the \idf{}. We use \texttt{numactl} pinning to control on which socket the Spark executor can allocate threads and memory. Figure~\ref{fig:deployment_analysis} summarizes our findings. We plot five combinations of executors and cores per executor, and {\sc numa} pinning. For this experiment we ran a \emph{join} operation between the 1B edge table of SNB SF-1000 and a 10M subset of it (XL join size in Table~\ref{tab:join_sizes}). More fine-grained executors perform better, and {\sc numa} pinning is able to further reduce the running time. In all subsequent experiments (with the exception of the ones performed in the cloud) we use the best performing configuration from Figure~\ref{fig:deployment_analysis}. On dual-socket machines, each of which having 8 cores per socket, we start 4 executors per machine -- two per NUMA domain. Each executor is allocated 4 cores.

\begin{figure}[t]
\centering
\includegraphics[width=0.8\columnwidth]{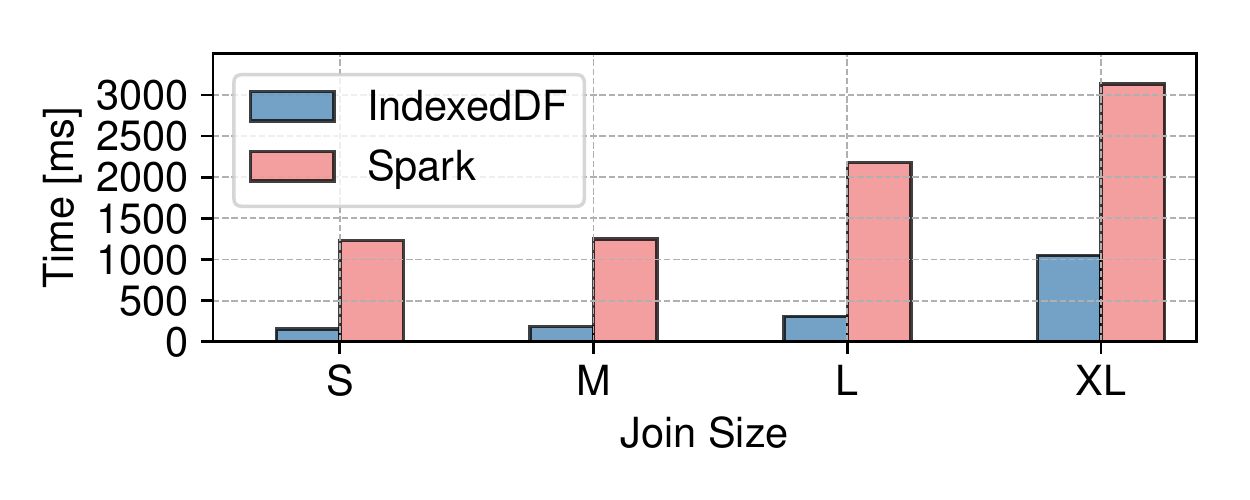}
\vspace*{-0.6cm}
\caption{The performance of the \idfregular{} versus vanilla Spark for different sizes of the join probe relation (see Table~\ref{tab:join_sizes}). The bars represent the average runtimes over 10 repetitions.}
\label{fig:indexed_join}
\vspace*{-0.5cm}
\end{figure}

\textbf{Row Batch Size.} Another important low-level parameter for the \idfregular{} is the row batch \emph{size} (i.e., the granularity of the buffers used to store data for each partition of the \idfregular{}). Larger allocations determine less batches per partition. We performed a similar experiment to the one described above to perform \emph{reads} (i.e., joins) and measured append performance to determine \emph{write performance}. Figure~\ref{fig:bs_sensitivity} summarizes the findings of the experiment. We normalize the performance for both reads and writes by the performance achieved using 4\,KB batches, the default OS page size. We identify a sweet-spot for both read and write performance at batch sizes of 4\,MB. We experimented also with larger batch sizes, of up to 128\,MB, but these perform exceptionally poorly for writes and are excluded from the figure. 

\subsection{Scalability Analysis}\label{subsec:scalability}

Spark is designed to scale-out on compute clusters with many machines. To reduce the network communication~\cite{DBLP:conf/eurosys/ZahariaBSESS10} Spark employs delay scheduling, a technique that aims at co-locating the tasks with their data. Incorporating indexing in Spark may, however, impact the scalability of the applications. We evaluate the scalability properties of \emph{join} operations on the \idfregular{} using the SNB benchmark by varying both the cluster size and the input data size. 

\textbf{Horizontal and Vertical.} 
To showcase the scalability properties of the \idf{}, we use a join between the 1B edge table of SNB SF-1000 and a 10M subset of it (XL join size in Table~\ref{tab:join_sizes}). We show that the \idf{} scales well both \emph{horizontally} and \emph{vertically} for such join queries in which our indexing structure requires a non-trivial amount of shuffle operations.

\begin{figure}[t]
\centering
\includegraphics[width=0.85\columnwidth]{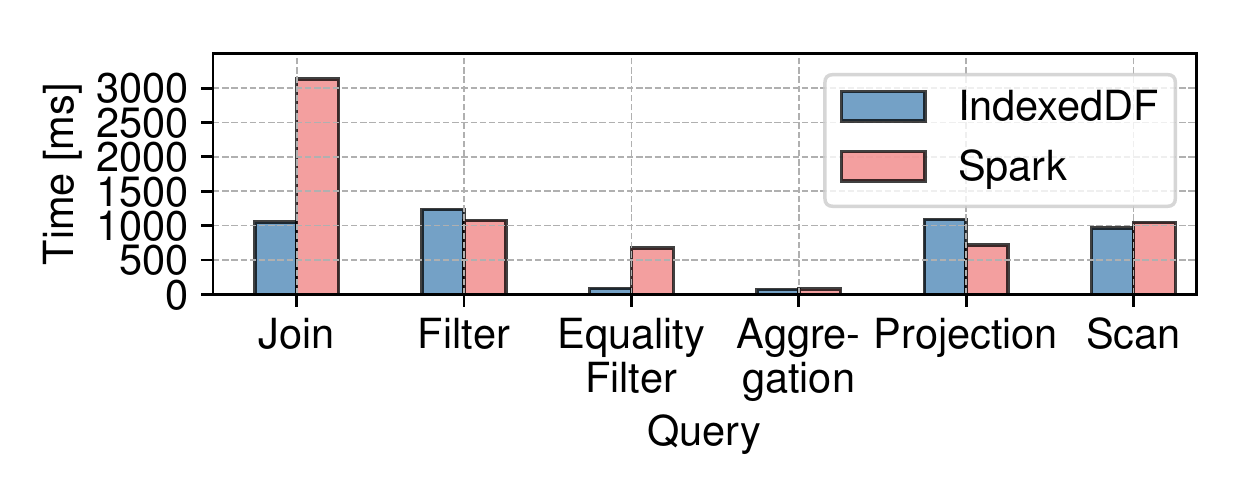}
\vspace*{-0.6cm}
\caption{The performance of the \idf{} versus vanilla Spark for different SQL operators. The bars depict the average runtimes over 10 repetitions.}
\label{fig:indexed_microbench}
\vspace*{-0.5cm}
\end{figure}

Figure~\ref{fig:horizscal} depicts the horizontal scalability behavior of the \idfregular{}. We increase the number of worker machines in our Spark cluster from 2 to 32 and we keep the input dataset size constant. We observe that the speed-up is sub-linear because increasing the cluster size results in more network communication. However, the same behavior applies for regular Spark operation.
Furthermore, in Figure~\ref{fig:vertscal} we show how our \idfregular{} scales vertically. In this experiment, we setup a cluster with 4 worker machines and we vary the number of cores used by the Spark executors to run tasks from 1 to 16. For simplicity, in this experiment we configured a single executor per worker machine. We find that the \idfregular{} has close to linear scaling with the number of cores per executor. 

\textbf{Dataset Size.} We compare the operation of a hash-join in Spark with our \idfregular{}. When two tables are hash-joined, Spark first creates a hash-table on the smaller side of the join which is called the {\it build} relation. When this relation is small (less than 10\,MB), Spark broadcasts it across all worker machines in the cluster to co-locate it with the opposite table of the join -- the probe relation. For each row in the {\it probe} relation, Spark finds corresponding rows from the build relation by lookups in the hash table.

The operation of the \idfregular{} is in contrast with the execution of a typical Spark hash-join. The index is always pre-built on the side of the join that remains in place, i.e., the larger table (the build side). The index is then used for locally probing the other table, whose partitions are shuffled over the network to co-locate them with the index. When the probing relation increases, the networked communication is prone to become a bottleneck. In Table~\ref{tab:join_sizes} we depict the probe relation sizes used in our join queries. We find that irrespective the probe size, our \idfregular{} is faster than Spark with speed-ups in the range of 3 and 8 (Figure~\ref{fig:indexed_join}).

\begin{figure}[t]
    \centering
    \includegraphics[width=0.85\columnwidth]{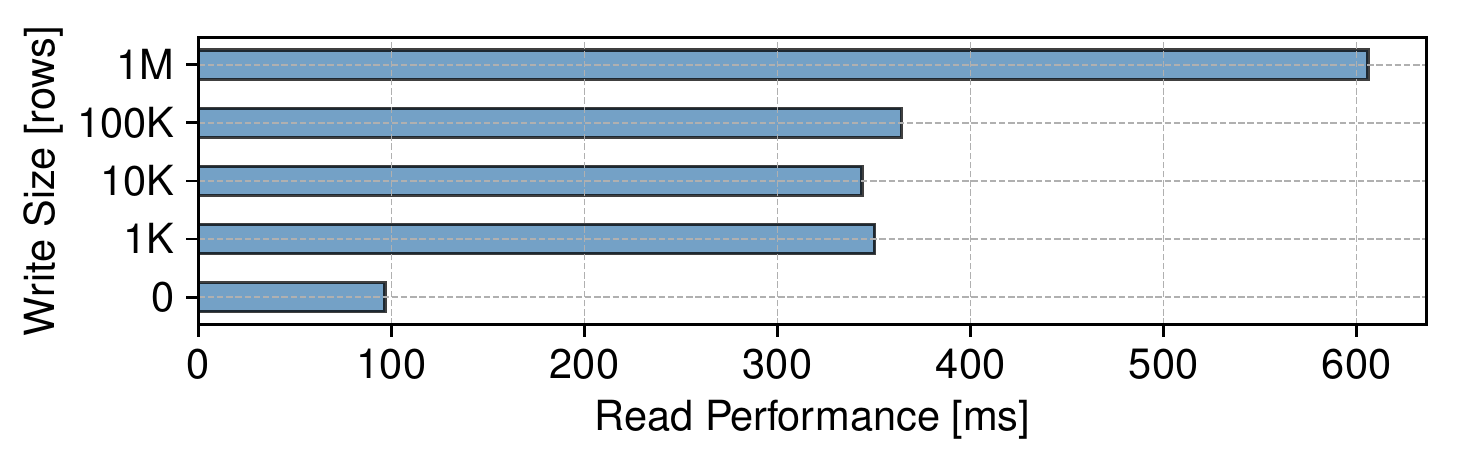}
    \vspace*{-0.4cm}
    \caption{Read performance latency increase when writing various amounts of rows at a time. Experiments are performed 200 times and results shown represent the mean.}
    \label{fig:read_latency}
    \vspace*{-0.4cm}
\end{figure}

\subsection{\idfregular{} Microbenchmarks}\label{subsec:micro}
The Spark dataframe and SQL APIs contain a plethora of operations that programmers can employ to manipulate large-scale datasets. We want to ensure that the \idfregular{} delivers similar performance with vanilla Spark on such operations as well as negligible storage overhead.

\begin{figure}[t]
    \centering
    \includegraphics[width=0.85\columnwidth]{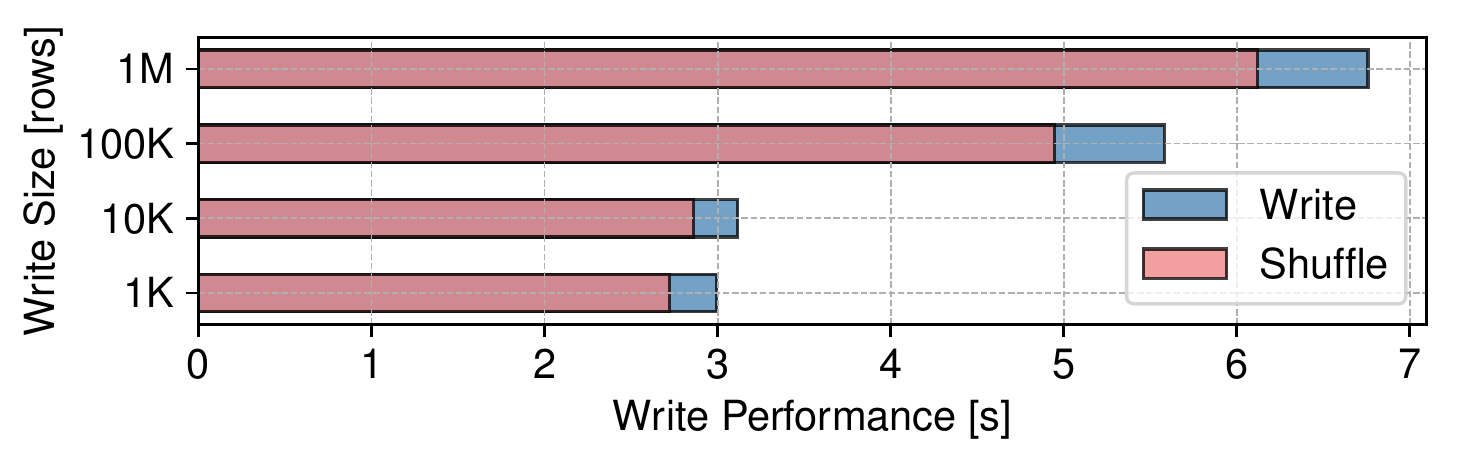}
    \vspace*{-0.4cm}
    \caption{Write performance throughput of the \idfregular{} for  various amounts of rows written at a time. Experiments are performed 200 times and the performance results are cumulated. This result applies to both \emph{appendRows} and \emph{createIndex} since they are backed by the same writing mechanism.}
    \label{fig:write_tput}
    \vspace*{-0.4cm}
\end{figure}

\textbf{Microbenchmarks.} We compare the performance of the \idfregular{} versus vanilla Spark on a selection of SQL operators including join, filter, projection, aggregation, and scan using an in-memory input dataframe that incorporates the large 1\,B rows edge table of the SNB benchmark (see Table~\ref{tab:join_sizes}).
Since the join and filtering operators naturally use the index their performance is significantly improved by the \idfregular{} as we show in Figure~\ref{fig:indexed_microbench}. We observe that the projection and non-equality filters are the only operators that suffer slowdowns because of our \idfregular{}. This is because our in-memory representation of the data is based on a row structure which is less efficient than the columnar format adopted by the Spark cache for projections or analyzing single columns.

\textbf{Append \& Index Creation.} Writing data into the \idfregular{} has two components: the \emph{latency} it adds to read operations (due to the extra row materialization and extra shuffles when actually writing the data), and the \emph{throughput} of effectively appending the data. To measure these, we performed two experiments. The first entails running 200 S join operations (see Table~\ref{tab:join_sizes}) in a sequence. Every 5 join operations we also issue an append. This models real-world behavior of users who need to query data sources that get written into regularly. The read performance is influenced by the size of the append. Figure~\ref{fig:read_latency} shows the results we achieved. The read performance is significantly influenced by the write size. Writes of at most 100K rows slow down reads by a factor of 3X, but larger writes double the latency to a factor of 6X. However, all this is still acceptable compared to regular Spark operation, where all join sizes take longer than 1 second (see Figure~\ref{fig:indexed_join}), without tolerating appends.

To measure effective writing throughput we performed 200 appends of various row sizes. Figure~\ref{fig:write_tput} presents the results. It is immediate that most of the \emph{write} time is dominated by shuffles. This is expected, because the rows have to be sent to the \idfregular{} partitions that are responsible for storing them, as the data is hash-partitioned. Spark is known to not be efficient for networked operations~\cite{DBLP:conf/eurosys/ZhangCSCF18}. All possible write implementations on Spark would suffer the same shuffle penalty. It is important to notice that the results are similar for both \emph{append} and \emph{createIndex}, as the two APIs perform the same internal operations. Note that the results presented in Figure~\ref{fig:write_tput} are cumulated performance results over 200 executions. In the topmost experiment we inserted 200\,M rows in batches of 1\,M rows, which took just below 7 seconds.

\begin{figure}[t]
    \centering
    \includegraphics[width=0.85\columnwidth]{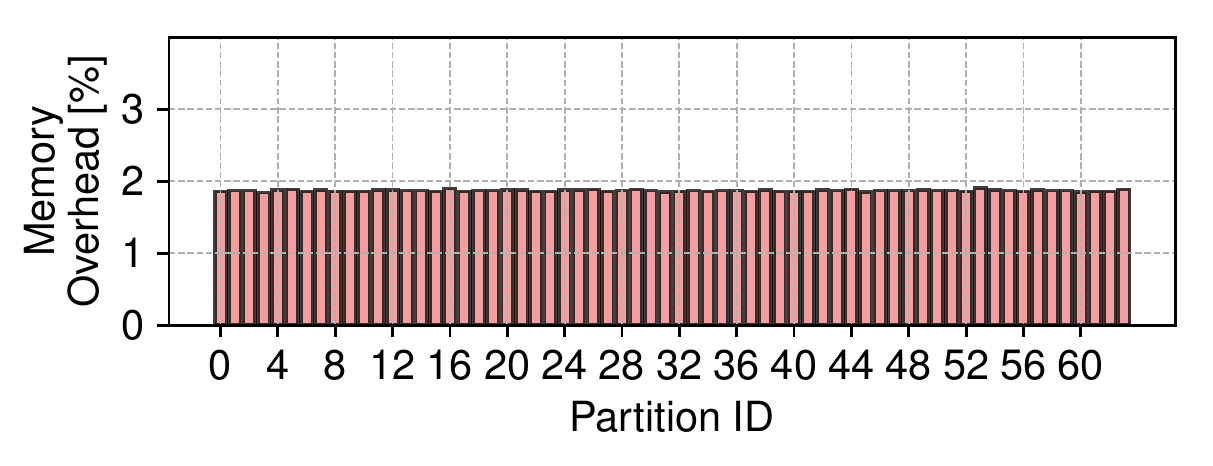}
    \vspace*{-0.4cm}
    \caption{The memory overhead of the \idfregular{} for each partition of a 30~GB table.}
    \label{fig:mem_overhead}
    \vspace*{-0.4cm}
\end{figure}

\textbf{Memory Overhead.} Storing a tree-based index next to the actual data may result in memory overhead. To assess how large this overhead is we investigated in detail our previous experiment. When creating the index, the 30\,GB SNB edge table is split into partitions, each Spark node storing a subset of those partitions. In our design, each partition stores its own local index in the Indexed Batch RDD, as described in Section~\ref{sec:design}. We measure the overhead of the cTrie index for each of these partitions. To this end, we instrument the \idfregular{} code with the JAMM memory meter~\cite{jamm}. Figure~\ref{fig:mem_overhead} plots the memory overhead for the 64 partitions of the SNB edge table. We find that the memory overhead for the \idfregular{} is consistently lower than $2\%$ and therefore negligible in comparison with the performance benefits achieved by indexing. 

\begin{figure}[t]
    \centering
    \includegraphics[width=0.85\columnwidth]{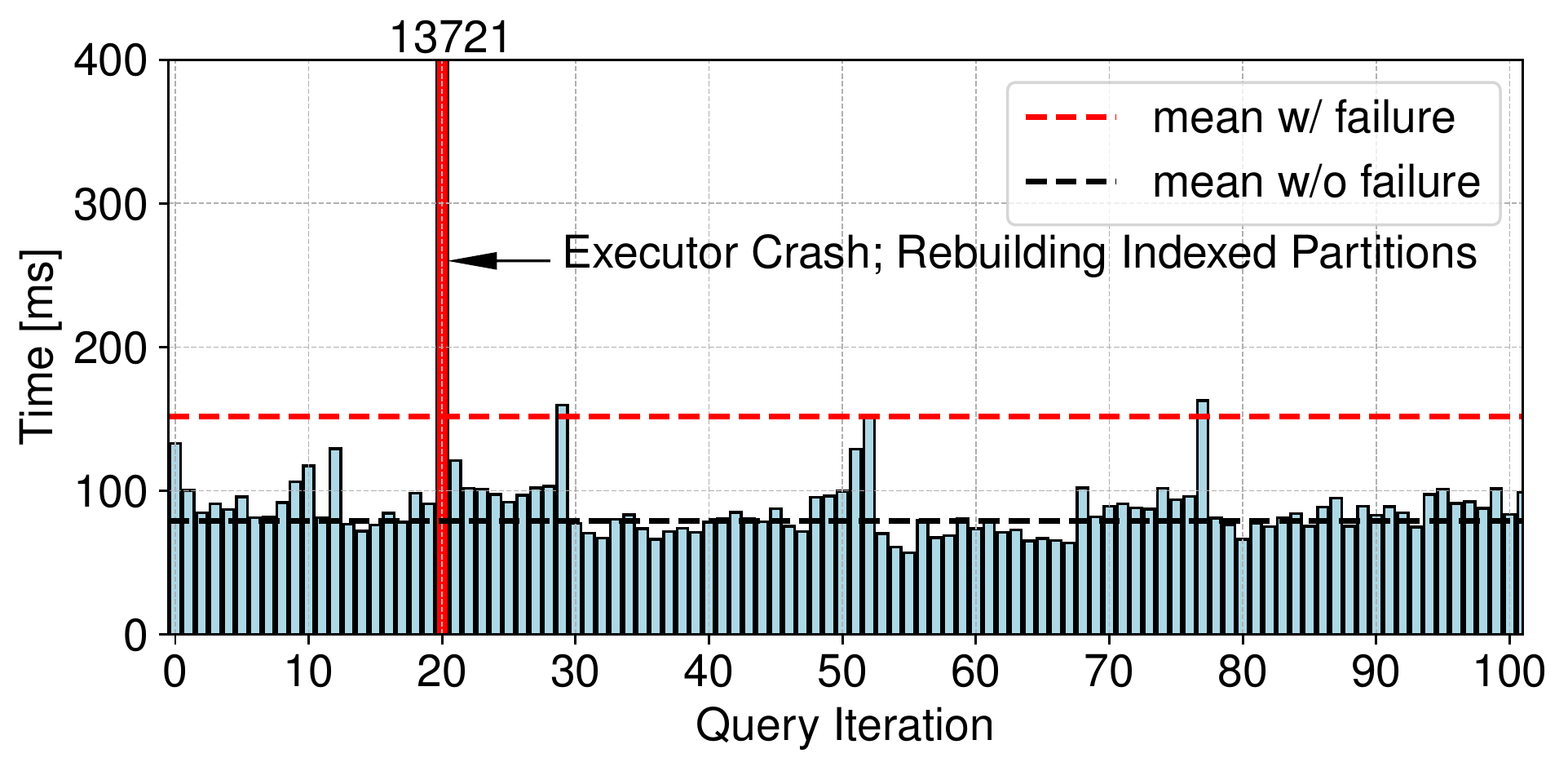}
    \vspace*{-0.4cm}
    \caption{\idfregular{} fault-tolerance overhead when one Spark executor fails during the execution of 200 join queries. The horizontal axis ends at 100 for visibility.}
    \label{fig:ft-overhead}
    \vspace*{-0.51cm}
\end{figure}

\textbf{Fault-Tolerance.} In the Spark ecosystem, fault-tolerance is achieved via re-computation based on the lineage graph. Whenever a node is lost, all its data has to be re-created by re-running the operations that led to its creation. For the \idf{} this mechanism is no different, however, it entails more work (Section~\ref{sec:ft}). If an indexed partition is lost, then the index has to be re-created, and if there were any appends on that particular partition, these have to be replayed as well. This process adds an overhead, but we show that once this overhead is paid for, the \idfregular{} performance falls back to normal. 
In a cluster of 8 nodes, we performed 200 S join operations (see Table~\ref{tab:join_sizes}) and during the execution time we manually killed a Spark executor which was holding 4 indexed partitions. Figure~\ref{fig:ft-overhead} shows the results. We notice that the failure took place during the execution of the 20th query. Re-creating the index extends the execution time of this query to over 13s, but subsequent queries operate at regular speed and the average execution time is only increased marginally. In conclusion, the overhead incurred by re-creating the index can be tolerated in typical environments where failures occur but are infrequent. 

\subsection{\idfregular{} in the Real-World}\label{subsec:realworld}
In this section, we analyze the performance of the \idfregular{} using real-world workloads and datasets, as well as production deployments. We analyze the performance of a set of queries selected from the real-world SNB benchmark as well as the state-of-the-art TPC-DS and US Flights workloads.

\begin{figure}[t]
\centering
\includegraphics[width=0.85\columnwidth]{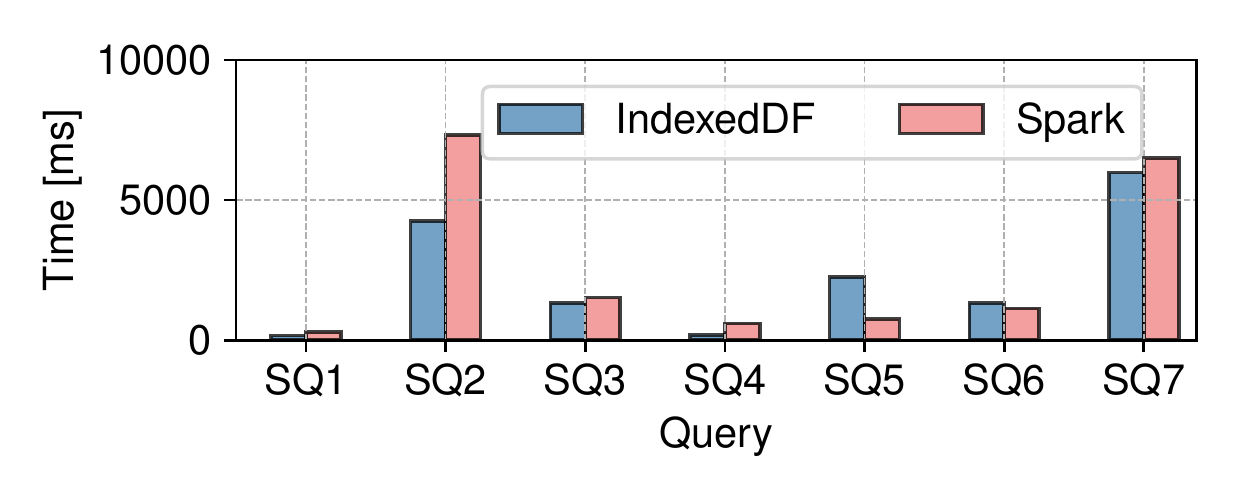}
\vspace*{-0.55cm}
\caption{The improvement of the SNB queries on \idf{} versus Spark on an input dataset of SF 300.}
\label{fig:indexed_SNB}
\vspace*{-0.4cm}
\end{figure}

\textbf{Social-Network Benchmark.} We run the complete set of short read SNB queries on an input dataset of scale factor 300. The results are presented in Figure~\ref{fig:indexed_SNB}. We observe that the \idf{} speeds up all queries, with the exception of SQ5 and SQ6, which are unable to use the index properly. This behavior is similar to the issue identified in Subsection~\ref{subsec:micro} in which we found that a \emph{columnar} data representation performs better than a \emph{row-based} representation for projections. The access patterns of these two queries trigger the inefficiencies of the row-based representation. 
However, data representation is orthogonal to the design of the \idfregular{} and column-based solutions may affect the scan performance because they are likely to produce more cache misses when materializing all rows~\cite{DBLP:journals/tos/WenLLDH19}.


\begin{figure}[t]
    \centering
    \includegraphics[width=0.85\columnwidth]{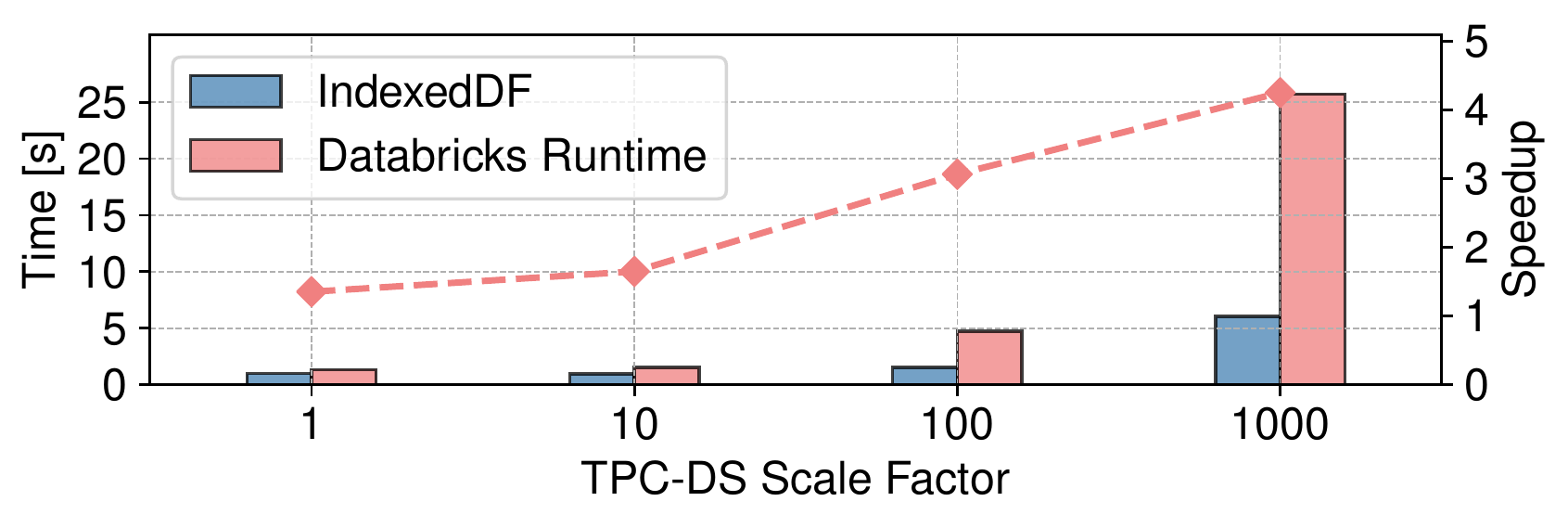}
    \vspace*{-0.4cm}
    \caption{Performance of the \idfregular{} relative to Databricks Runtime using TPC-DS dataset and a join query running on 16 i3.8xlarge Amazon EC2. Bars represent average results computed over 10 runs. Secondary vertical axis represents speedup.}
    \label{fig:dbr_tpcds}
    \vspace*{-0.45cm}
\end{figure}

\textbf{Integration in Production.} The \idfregular{} is implemented as a lightweight library, which can be included in any existing Spark programs, with minimal modifications that basically require attaching the library to the Spark cluster and invoking a simple API to create the index. In this way, we integrated our library in the Databricks Runtime 
to speed-up the performance of join and look-up queries. In this benchmark we use the BroadcastHashJoin implementation, which is faster than the notoriously slow SortMerge Join.

\textbf{TPC-DS.} Figure~\ref{fig:dbr_tpcds} shows our results with TPC-DS datasets of increasing size when running a join query (see Table~\ref{tab:queries_description}) on an Amazon EC2 cluster of 16 \texttt{i3.8xlarge} running the Databricks Runtime. We observe that by having the \idfregular{} created upfront we are able to significantly improve the performance by large margins. We tested several orders of magnitude of the TPC-DS scale factor, from 1 to 1000. The trend here is clear: the larger the dataset, the larger the gap between the indexed version of the join compared to its non-indexed version. This is an effect of the large amount of data that can be filtered out by using the index: the larger the dataset size, the more data is filtered out by the index.

\begin{figure}[t]
    \centering
    \includegraphics[width=0.85\columnwidth]{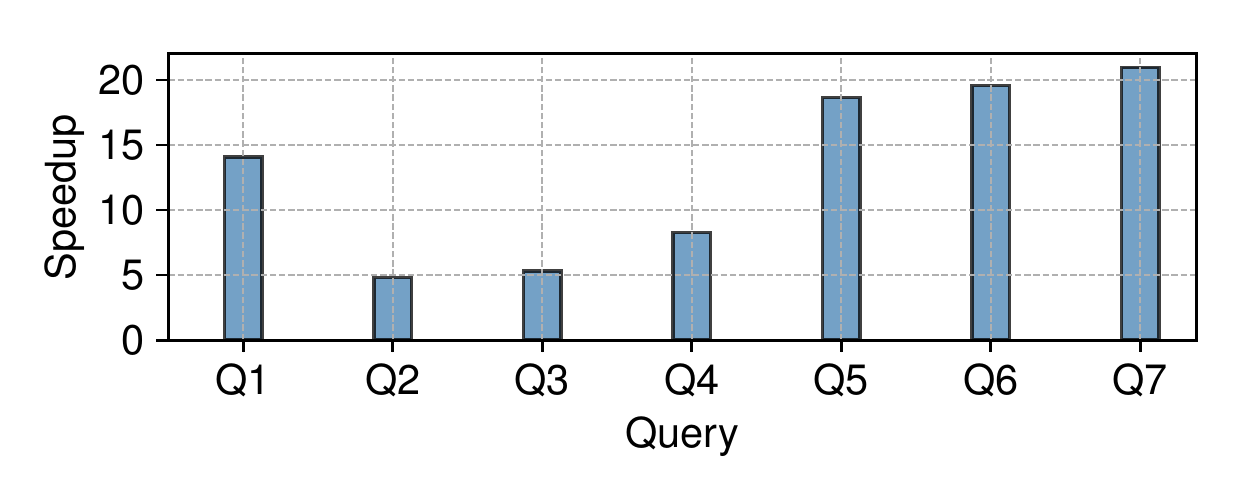}
    \vspace*{-0.6cm}
    \caption{Speedup of the \idfregular{} compared to the Databricks Runtime on 7 queries running on the US Flights dataset. Bars represent average speedup over 10 runs.}
    \label{fig:db_flights}
    \vspace*{-0.4cm}
\end{figure}

\textbf{US Flights.} Figure~\ref{fig:db_flights} shows our results on the Databricks Runtime when executing several queries analyzing the US Flights dataset (see Table~\ref{tab:queries_description}). We compare the performance of the \idfregular{} operations on both string and integer columns. Our main findings are as follows. As expected, \idfregular{} clearly outperforms the stock Databricks Runtime, by factors of 5-20X. The largest speedups are achieved for integer-key based filters (Q5--Q7). The string-based filter (Q2) is sped up only up to 5X because non-primitive type data incurs additional overhead to be used as a key. Strings need to be hashed into an 32-bit number which is then used as a key in the cTrie. For similar reasons, integer-key joins are sped up more than the string-key joins.

\section{Discussion and Related Work}

Indexing has been successfully and extensively used in database design for decades to accelerate many types of database operations, including joins~\cite{DBLP:journals/tods/Roussopoulos82,DBLP:journals/tods/Valduriez87,DBLP:journals/sigmod/ONeilG95}. These techniques were not only applicable in single-machine setups, but also in distributed databases~\cite{DBLP:conf/sigir/DanzigANO91,DBLP:conf/ccgrid/NamS05,DBLP:conf/bigdataconf/DehneKRZZ13}, in both OLTP~\cite{DBLP:conf/sigmod/DiaconuFILMSVZ13} and OLAP~\cite{DBLP:conf/icde/GuptaHRU97} systems. These database systems that support indexing have in common an ability to handle fine-grained writes, sometimes with strong transaction semantics.

Going forward, in the \emph{big data era} the major technology switch was toward MapReduce-like systems~\cite{DBLP:conf/osdi/DeanG04}, which are designed for achieving performance through scaling out and simplicity. In this category, we can include not only Hadoop~\cite{DBLP:books/daglib/0022835}, but also Spark~\cite{zaharia2012resilient}, DryadLinq~\cite{DBLP:conf/osdi/YuIFBEGC08}, or Naiad~\cite{DBLP:conf/sosp/MurrayMIIBA13}, which extend the simplistic MapReduce model to more complex dataflows. Such big data processing systems are thought to be complementary to parallel databases: Stonebraker et al. \cite{DBLP:journals/cacm/StonebrakerADMPPR10} consider the former to excel at complex analytics and ETL, while the latter at efficiently query-ing large datasets. One of the inefficiencies of MapReduce-like systems come from the fact that they are not designed to support indexing~\cite{dewitt2008mapreduce}.

Significant research effort has been invested into adding indexes to improve the performance of data processing systems. Hadoop has been updated with adaptive HDFS indexes~\cite{DBLP:journals/vldb/0007QSD14}, indexes at split level~\cite{DBLP:conf/edbt/EltabakhOSHPV13}. Moreover, indexes have been used to speed up Hive queries~\cite{DBLP:conf/bigdataconf/MofidpoorSR13} running on top of Hadoop. Indexes have also been built in Spark, for the purpose of quickly solving geospatial queries~\cite{xie2016simba,cui2017indexing,tang2016locationspark}. However, the downside of all such indexing techniques on current data processing systems is that they do not support any kind of fine-grained updates, either in place or appends, thus not being able to offer interactive performance for applications that do need updates.

SnappyData~\cite{DBLP:conf/cidr/MozafariRMMCBB17} supports finer-grained updates and indexes by integrating Spark with an external key-value store--Apache GemFire. Managing an external system next to Spark is an overhead that not many organizations can afford. Furthermore, existing Spark programs need be significantly modified to support SnappyData. In our experiments, we were unable to get consistent results from SnappyData when trying to run the benchmarks used in our evaluation. Koalas~\cite{koalas} is another indexed processing system that uses Spark as a backend. Due to this design, its performance is bottlenecked by the behavior of default Spark when it comes to fine-granuar updates as exhibited by our benchmarks. In our experiments, Koalas was orders of magnitude slower for both indexed joins and point queries. Dask is a python-based distributed framework that offers Pandas-like functionality. However, language overheads like global interpreter locks caused Dask in our experiments to be orders of magnitude slower than the \idfregular{}. 



Another kind of large-scale storage systems for big data processing that provide indexes are the so-called cloud data lakes, such as Delta Lake~\cite{armbrust2020delta} and Helios~\cite{potharaju2020helios} or Hyperspace~\cite{potharajuhyperspace}. They offer a tightly-coupled integration between analytics engines and large-scale cloud (object) storage systems. In contrast to these systems, which offer secondary storage indexing, our work could be seen as an in-memory indexed cache, which is easy to integrate even with systems such as Delta Lake or Helios.

The Timely Dataflow system~\cite{timely} based on the Naiad~\cite{DBLP:conf/sosp/MurrayMIIBA13} design promises to solve most sources of inefficiencies in current data processing systems. It is unclear at the moment whether such systems would even benefit from indexes, but their current performance promises to deliver beyond the inefficiencies of current data processing systems. 
However, their application in practice lags behind the popularity of Spark. Finally, using machine learning techniques to learn and optimize indexes~\cite{DBLP:journals/corr/abs-1712-01208} is a promising research direction, provided it can build upon robust indexing mechanisms. \idfregular{}s could be such a mechanism for AI-driven~indexes
\section{Conclusion} \label{sec:concl}

In this paper we argue for bridging the divide between traditional relational database systems and data analytics platforms like Spark to get the best of both worlds. Concretely, we showed that adding indexing to the main underlying abstractions of Spark can greatly improve performance in many cases. The presented \idfregular{} brings indexing to the Spark data processing engine, as well as fine-grained appends with minimum performance and memory overheads, and large-scale gains for workloads that make use of the index, such as joins or point lookups. The \idfregular{} is designed as a lightweight library that can be included in any Spark program, from any Spark distribution--be it Apache, or Databricks Runtime--to achieve a boost in performance for workloads that make use of indexes, of up to 20X. 

\section*{Acknowledgements}
Part of this work was conducted while the first author was an intern at Databricks. We would like to thank Herman van Hovell, Adrian Ionescu for their suggestions on the implementation of the project, as well as Matei Zaharia for his valuable comments on the manuscript of the paper. The work in this article was in part supported by The Dutch National Science Foundation NWO Veni grant VI.202.195.



\bibliographystyle{IEEEtran}
\bibliography{references}

\vskip -2.5\baselineskip plus -1fil

\end{document}